\documentclass[10pt,conference]{IEEEtran}

\usepackage{amsfonts}
\usepackage[dvips]{graphicx}
\usepackage{times}
\usepackage{cite}
\usepackage{amsmath}
\usepackage{cases}
\usepackage{array}
\usepackage{dsfont}
\usepackage{amssymb}

\usepackage{stfloats}
\usepackage{slashbox}
\usepackage{graphicx}
\usepackage{footnote}
\usepackage{booktabs}
\usepackage{array}
\usepackage{bm}
\usepackage{multicol}

\begin{document}

\title{Generalized Signal Alignment For MIMO Two-Way X Relay Channels}
\author{\IEEEauthorblockN{Kangqi Liu, Meixia Tao, Zhengzheng Xiang and Xin Long}
\IEEEauthorblockA{Dept. of Electronic Engineering, Shanghai Jiao Tong University,  Shanghai, China\\
Emails: \{forever229272129, mxtao, 7222838, lx\_yokumen\}@sjtu.edu.cn}
}

\maketitle

\begin{abstract}
We study the degrees of freedom (DoF) of MIMO two-way X relay channels. Previous work studied the case $N < 2M$, where $N$ and $M$ denote the number of antennas at the relay and each source, respectively, and showed that the maximum DoF of $2N$ is achievable when $N \leq \lfloor\frac{8M}{5}\rfloor$  by applying signal alignment (SA) for network coding and interference cancelation. This work considers the case $N>2M$ where the performance is limited by the number of antennas at each source node and conventional SA is not feasible. We propose a \textit{generalized signal alignment} (GSA) based transmission scheme. The key is to let the signals to be exchanged between every source node align in a transformed subspace, rather than the direct subspace, at the relay so as to form network-coded signals. This is realized by jointly designing the precoding matrices at all source nodes and the processing matrix at the relay. Moreover, the aligned subspaces are orthogonal to each other. By applying the GSA, we show that the DoF upper bound $4M$ is achievable when $M \leq \lfloor\frac{2N}{5}\rfloor$ ($M$ is even) or $M \leq \lfloor\frac{2N-1}{5}\rfloor$ ($M$ is odd). Numerical results also demonstrate that our proposed transmission scheme is feasible and effective.
\end{abstract}

\renewcommand{\thefootnote}{\fnsymbol{footnote}}
\setcounter{footnote}{-1}
\footnote{This work is supported by the National 973 project under grant 2012CB316100 and by the NSF of China under grants 61322102 and 61329101.}

\section{Introduction}

Wireless relay has been an important ingredient in both ad hoc and infrastructure-based wireless networks. It shows great promises in power reduction, coverage extension and throughput enhancement. In the simplest scenario, a relay node only serves a single user. This forms the classic \textit{one-way relaying} and the relay strategies are maturing. With the rapid expansion of multi-user communications, a relay has become very much like a wireless gateway where multiple users communicate with each other via a common relay. A fundamental question that arises is what is the maximum number of data streams that can be transmitted and how to achieve it. This leads to the analysis of degrees of freedom (DoF) and also drives the development of more advanced relay strategies for efficient relay-assisted multi-user communication.

The recently proposed \textit{two-way relaying} is such an advanced relay method that offers high spectral efficiency in a system where two users exchange information with each other through a relay \cite{Rankov}. The key idea is to apply physical layer network coding (PLNC) at the relay end \cite{ZhangSL} \cite{Katti}. With PLNC, the maximum achievable DoF of the MIMO two-way relay channel is $2\min\{M,N\}$ \cite{Vaze}, where $M$ and $N$ denote the number of antennas at each source node and the relay, respectively. When there are three or more users exchanging information with each other via a common relay, PLNC is not enough to achieve the DoF of the network.

Based on the idea of interference alignment\cite{Jafar} \cite{Maddah}, another promising technique, \textit{signal alignment} (SA) is firstly proposed in \cite{Lee2} to analyze the maximum achievable DoF for MIMO Y channel, where three users exchange independent messages with each other via a relay. By jointly designing the precoders at each source node, SA is able to align the signals from two different source nodes in a same subspace of the relay node. By doing so, the two data streams to be exchanged between a pair of source codes are combined into one network-coded symbol and thus the relay can forward more data streams simultaneously.
It is proved that with SA for network-coding and network-coding aware interference nulling the theoretical upper bound $3M$ of DoF is achievable when $N \geq \lceil \frac{3M}{2}\rceil$ \cite{Lee1}. Here, again, $M$ and $N$ denote the number of antennas at each source node and the relay node, respectively. The extension to K-user MIMO Y Channels is considered in \cite{Lee}.

In \cite{Xiang}, SA is applied in MIMO two-way X relay channel, where there are two groups of source nodes and one relay node, and each of the two source nodes in one group exchange independent messages with the two source nodes in the other group via the relay node. It is shown that the DoF upper bound is $2\min\{2M,N\}$, and the upper bound $2N$ is achievable when $N \leq \lfloor \frac{8M}{5} \rfloor$ by applying SA and interference cancelation.

In this paper, we are interested in the ability of the DoF upper bound $4M$ for MIMO two-way X relay channel in the case $N>2M$, where the performance is limited only by the number of antennas at each source. It is worth mentioning that SA is not feasible under the antenna configuration $N \geq 2M$. The reason is as follows. Recall that SA condition is

\begin{equation}\label{SA}
{\bf H}_{1,r}{\bf v}_{1}={\bf H}_{2,r}{\bf v}_{2},
\end{equation}
where ${\bf H}_i$ is an $N \times M$ matrix (corresponding to the channel matrix from source $i$ to relay) and ${\bf v}_i$ is an $M \times 1$ vector (corresponding to the beamforming vector of source $i$).  The above alignment condition can be rewritten as

\begin{equation}\label{SA_1}
\left[{\bf H}_{1,r}~~ -{\bf H}_{2,r} \right]
\left[
\begin{array}{ccc}
{\bf v}_{1} \\
{\bf v}_{2} \\
\end{array}
\right]
=0.
\end{equation}
Clearly, for \eqref{SA_1} to hold, one must $N<2M$, or equivalently $M > \frac{N}{2}$.

To achieve the maximum DoF at $M \le \frac{N}{2}$, we propose a new transmission scheme, named \textit{generalized signal alignment} (GSA). Compared with the existing SA, the proposed GSA has the following major difference. The signals to be exchanged do not align directly in the subspace observed by the relay. Instead, they are aligned in a transformed subspace after certain processing at the relay, which is orthogonal to each other. This is done by jointly designing the precoding matrices at the source nodes and the processing matrix at the relay node. With the proposed GSA, we show that the total DoF upper bound $4M$ of MIMO two-way X relay channel is achievable when $M \leq \lfloor\frac{2N}{5}\rfloor$($M$ is even) or $M \leq \lfloor\frac{2N-1}{5}\rfloor$($M$ is odd).

The remainder of the paper is organized as follows. In Section II, we introduce the system model of the MIMO two-way X relay channel. In Section III, we introduce the GSA transmission scheme with a motivate example. In Section IV, we analyze the achievability of the DoF upper bound when $M<\frac{N}{2}$. In Section V, we show our numerical results. Section VI presents concluding remarks.

Notations: $(\cdot)^{T}$, $(\cdot)^{H}$ and  $(\cdot)^{\dag}$ denote the transpose, Hermitian transpose and the Moore-Penrose pseudoinverse, respectively. $\varepsilon[\cdot]$ stands for expectation. $|{\bf x}|$ means 2-norm of vector ${\bf x}$. $\mbox{span} ({\bf H})$ and ${\mbox{null} ({\bf H})}$ stand for the column space and the null space of the matrix ${\bf H}$, respectively. $\mbox{dim}({\bf H})$ denotes the dimension of the column space of ${\bf H}$. $\langle{\bf x}\rangle$ denotes normalization operation on vector ${\bf x}$, i.e.$ \langle {\bf x}\rangle= \frac{\bf x}{|\bf x|}$. $\lfloor x \rfloor$ denotes the largest integer no greater than $x$. $\lceil x \rceil$ denotes the smallest integer no less than $x$.
{\bf I} is the identity matrix.

\section{System Model}

We consider the same MIMO two-way X relay channel as in \cite{Xiang} and shown in Fig.~1. It consists of two groups of source nodes, each equipped with $M$ antennas, and one relay node, equipped with $N$ antennas. Each source node exchanges independent messages with each source node in the other group with the help of the relay. The independent message transmitted from source $i$ to source $j$ is denoted as $W_{ij}$. At each time slot, the message is encoded into a $d_{ij} \times 1$ symbol vector ${\bf s}_{ij}$, where $d_{ij}$ denotes the number of independent data flows from source $i$ to source $j$.

Taking source node $1$ for example, the transmitted signal vector ${\bf x}_1$ from source node $1$ is given by
\begin{eqnarray}\label{x_1}
{\bf x}_1 ={\bf V}_{13}{\bf s}_{13} + {\bf V}_{14}{\bf s}_{14}= {\bf V}_1{\bf s}_1,
\end{eqnarray}
where $\textbf{s}_1=[\textbf{s}_{13},\textbf{s}_{14}]^T$, $\textbf{V}_1= [\textbf{V}_{13}, \textbf{V}_{14}]$, $\textbf{V}_{13}$ and $\textbf{V}_{14}$ are the $M \times d_{13}$ and $M \times d_{14}$ precoding matrices for the information symbols to be sent to source node 3 and 4, respectively.

The communication of the total messages takes place in two phases: the multiple access (MAC) phase and the broadcast (BC) phase. In the MAC phase, all four source nodes transmit their signals to the relay. The received signal ${\bf y}_r$ at the relay is given by
\begin{eqnarray}\label{y_r}
{\bf y}_r=\sum\limits_{i=1}^{4}{\bf H}_{i,r}{\bf x}_{i}+{\bf n}_r,
\end{eqnarray}
where ${\bf H}_{i,r}$ denotes the frequency-flat quasi-static $N \times M$ complex channel matrix from source node $i$ to the relay and ${\bf n}_r$ denotes the $N\times 1$ additive white Gaussian noise (AWGN) with variance $\sigma_n^2$. The entries of the channel matrix ${\bf H}_{i,r}$ and those of the noise vector ${\bf n}_r$, are independent and identically distributed (i.i.d.) zero-mean complex Gaussian random variables with unit variance. Thus, each channel matrix is of full rank with probability $1$.

Upon receiving ${\bf y}_r$ in \eqref{y_r}, the relay processes these messages to obtain a mixed signal $\textbf{x}_r$,  and broadcasts to all the users. The received signal at source node $i$ can be written as
\begin{eqnarray}\label{y_i}
{\bf y}_i=\textbf{G}_{r,i}\textbf{x}_r+{\bf n}_i,
\end{eqnarray}
where ${\bf G}_{r,i}$ denotes the frequency-flat quasi-static $M \times N$ complex channel matrix from relay to the source node $i$, and $\textbf{n}_i$ denotes the AWGN at the node $i$. Each user tries to obtain its desirable signal from its received signal using its own transmit signal as side information.

\begin{figure}[t]
\begin{centering}
\includegraphics[scale=0.7]{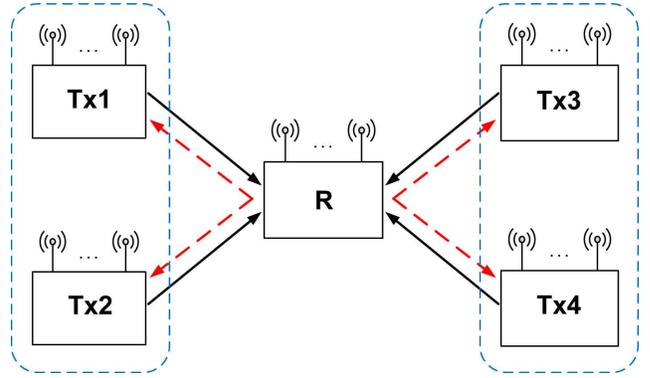}
\vspace{-0.1cm}
 \caption{MIMO Two-way X Relay Channel.}
\end{centering}
\vspace{-0.3cm}
\end{figure}


%

\section{Generalized Signal Alignment}
In this section, we shall introduce the GSA-based transmission scheme for MIMO two-way X relay channel when $M<\frac{N}{2}$. Note that for this case, the corresponding DoF upper bound is $4M$ \cite{Xiang}.

As a motivating example, we consider a system where each source has $M=2$ antennas and the relay has $N=5$ antennas. For this system, the proposed GSA-based transmission scheme achieves $d_{13}=d_{14}=d_{23}=d_{24}=d_{31}=d_{32}=d_{41}=d_{42}=1$, yielding a total DoF of $8$.

\subsection{MAC phase}
In the MAC phase, each source node transmits the precoded signals to the relay simultaneously.  Let source node $1$ transmits the data streams $s_{13}$ and $s_{14}$ by using beamforming vectors ${\bf v}_{13}$ and ${\bf v}_{14}$, respectively to source nodes $3$ and $4$. Similar notations are used for the other three nodes. Thus, there are totally $8$ data streams arriving at the relay. We rewrite the received signal \eqref{y_r} as
\begin{eqnarray}\nonumber
{\bf y}_r~~~~~~~~~~~~~~~~~~~~~~~~~~~~~~~~~~~~~~~~~~~~~~~~~~~~~~~~~~~~~~~\\\nonumber=\left[{\bf H}_{1,r}~{\bf H}_{2,r}~{\bf H}_{3,r}~{\bf H}_{4,r}\right]\left[\begin{array}{c}{\bf V}_1~~{\bf 0}~~{\bf 0}~~{\bf 0}\\{\bf 0}~~{\bf V}_2~~{\bf 0}~~{\bf 0}\\{\bf 0}~~{\bf 0}~~{\bf V}_3~~{\bf 0}\\{\bf 0}~~{\bf 0}~~{\bf 0}~~{\bf V}_4\end{array}\right]\left[\begin{array}{c}{\bf s}_1\\{\bf s}_2\\{\bf s}_3\\{\bf s}_4\end{array}\right]+{\bf n}_r\\\label{y_r_2_N}={\bf H}{\bf V}{\bf s}+{\bf n}_r,~~~~~~~~~~~~~~~~~~~~~~~~~~~~~~~~~~~~~~~~~~~~~~~~~~~~
\end{eqnarray}
where ${\bf H}$ is the $5\times 8$ overall channel matrix, ${\bf V}$ is the $8\times 8$ block-diagonal overall precoding matrix and $\bf s$ is the $8\times 1$ transmitted signal vector for all the source nodes, given by
\begin{equation}\label{s_3}
{\bf s}=\left[s_{13}~s_{14}~s_{23}~s_{24}~s_{31}~s_{32}~s_{41}~s_{42}\right]^T.
\end{equation}

Since the relay has only $5$ antennas, it is impossible for it to decode all the $8$ data streams. However, based on the idea of physical layer network coding, we only need to obtain the following network-coded symbol vector at the relay
\begin{eqnarray}
\textbf{s}_{\oplus}=\left[s_{13}+s_{31}, s_{14}+s_{41}, s_{23}+s_{32}, s_{24}+s_{42}\right]^T.
\end{eqnarray}

Next, we show that there exist a precoding matrix \textbf{V} for the source nodes and a processing matrix \textbf{A} for the relay node such that the network-coded symbol vector in \eqref{s_3} can be obtained. To state it formally, we introduce the following theorem.

\textit{Theorem 1}: Given the received signal model in \eqref{y_r_2_N}, there exists an $8\times8$ block-diagonal precoding matrix {\bf V} and a $4 \times 5$ relay processing matrix \textbf{A} such that
\begin{eqnarray}\nonumber\label{y_rrr}
\hat{{\bf y}}_r&=&{\bf A}{\bf y}_r\\\nonumber &=&{\bf A}{\bf H}{\bf V}{\textbf{s}}+{\bf A}{\bf n}_r\\&=&\textbf{s}_{\oplus}+{\bf A}{\bf n}_r.~~~~~~
\end{eqnarray}

\begin{proof}
Let $\textbf{a}_i$ denote the $i$-th row of \textbf{A}, for $i=1, \cdots, 4$. Each ${\bf a}_i$ can be thought as a combining vector for the transmitted signals of a source node pair. Specifically, we take $\textbf{a}_1$ for elaboration. We aim to design $\textbf{a}_1$ to align the transmitted signals from source pair (1,3) and cancel the undesired signals from source nodes 2 and 4. Thus, we design ${\bf a}_1$ such that it falls into the null space of ${\bf H}_{2,r}$ and ${\bf H}_{4,r}$
\begin{eqnarray}\label{a_1}
{\bf a}_1^T\subseteq \textbf{Null}~\big[{\bf H}_{2,r}~{\bf H}_{4,r}\big]^T.
\end{eqnarray}
Since $\big[{\bf H}_{2,r}~{\bf H}_{4,r}\big]^T$ is a $4\times 5$ matrix, we can always find such ${\bf a}_1$.
Similarly, the other rows of ${\bf A}$ can be obtained as:
\begin{eqnarray}
{\bf a}_2^T\subseteq \textbf{Null}~\big[{\bf H}_{2,r}~{\bf H}_{3,r}\big]^T.~\\\label{a_2}
{\bf a}_3^T\subseteq \textbf{Null}~\big[{\bf H}_{1,r}~{\bf H}_{4,r}\big]^T.~\\\label{a_3}
{\bf a}_4^T\subseteq \textbf{Null}~\big[{\bf H}_{1,r}~{\bf H}_{3,r}\big]^T.~\label{a_4}
\end{eqnarray}
where $\textbf{a}_2$ is for source pair (1,4), $\textbf{a}_3$ is for source pair (2,3) and $\textbf{a}_4$ is for source pair (2,4).


Given the processing matrix ${\bf A}$ at the relay, the effective channel in the MAC phase can be written as
\begin{equation}\label{C}
{\bf AH}=
\left[
\begin{array}{cccccccc}
  c_{11} & c_{12} & 0 & 0 & c_{15} & c_{16} & 0 & 0 \\
  c_{21} & c_{22} & 0 & 0 & 0 & 0 & c_{27} & c_{28} \\
  0 & 0 & c_{33} & c_{34} & c_{35} & c_{36} & 0 & 0 \\
  0 & 0 & c_{43} & c_{44} & 0 & 0 & c_{47} & c_{48}.
\end{array}
\right]
\end{equation}

Define $\textbf{C}_i,~i=1,2,3,4$ as
\begin{equation*}\label{C1}
\textbf{C}_1=
\left[
\begin{array}{cccccccc}
  c_{11} & c_{12} \\
  c_{21} & c_{22} \\
\end{array}
\right],~\textbf{C}_2=
\left[
\begin{array}{cccccccc}
  c_{33} & c_{34} \\
  c_{43} & c_{44} \\
\end{array}
\right],
\end{equation*}

\begin{equation}\label{C3}
\textbf{C}_3=
\left[
\begin{array}{cccccccc}
  c_{15} & c_{16} \\
  c_{35} & c_{36} \\
\end{array}
\right],~\textbf{C}_4=
\left[
\begin{array}{cccccccc}
  c_{27} & c_{28} \\
  c_{47} & c_{48} \\
\end{array}
\right].
\end{equation}
We can construct the precoding matrix for each source node as
\begin{eqnarray}\label{V_i}
{\bf V}_i={\bf C}_i^{-1},~i=1,2,3,4.
\end{eqnarray}

The above precoding matrix ${\bf V}_i$ can be seen as a zero-forcing based precoder to cancel inter-stream interference between those signals not to be exchanged for each source node pair. For example, ${\bf V}_1$ helps cancel the unwanted signal $s_{14}$ for source node pair $(1,3)$ and $s_{13}$ for source node pair $(1,4)$. Substituting \eqref{a_1} - \eqref{a_4} and \eqref{V_i} into \eqref{y_r_2_N}, we can obtain \eqref{y_rrr}. The theorem is thus proved.
\end{proof}

By jointly designing the precoding matrices at the source nodes and the processing matrix at the relay, our proposed GSA has successfully aligned the transmitted signals for each source node pair at the relay. The key steps of GSA are illustrated in Fig. $2$ and Fig. $3$. Fig. 2 shows that $\textbf{a}_1^T$ falls into the null space of $\textbf{H}_{2,r}$ and $\textbf{H}_{4,r}$. It is similar to other $\textbf{a}_i$ (i=2, 3, 4) by \eqref{a_1} - \eqref{a_4}. Fig. 3 shows the whole signal processing procedure to obtain the network-coded messages. Here, the signals received at relay after the effective channel $\textbf{AH}$ is firstly given. Then the signals are rotated by $\textbf{V}$ to be aligned in four orthogonal directions.

\begin{figure}[t]
\begin{centering}
\includegraphics[scale=0.8]{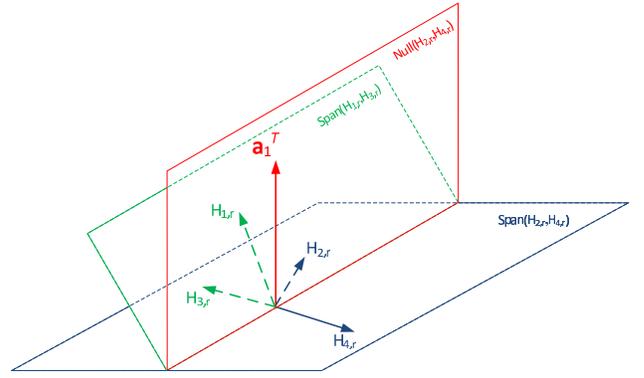}
\vspace{-0.1cm}
 \caption{The construction the processing matrix at relay.}
\end{centering}
\vspace{-0.3cm}
\end{figure}


\begin{figure*}[htbp!]
\begin{centering}
\includegraphics[scale=0.56]{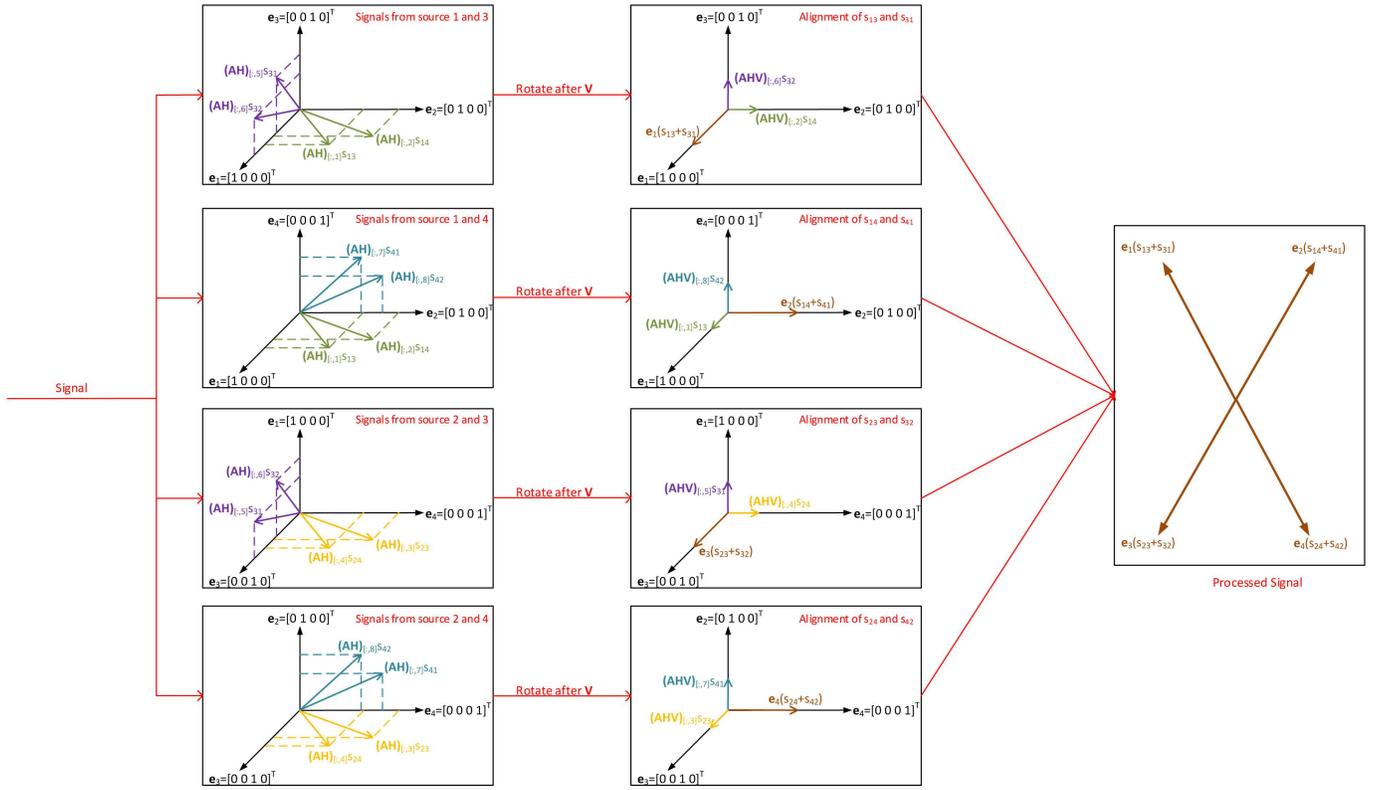}
\vspace{-0.1cm}
 \caption{Generalized signal alignment procedure.}
\end{centering}
\vspace{-0.3cm}
\end{figure*}


\subsection{BC Phase}
During the BC phase, the relay broadcasts an estimate of the four network coded symbols using the precoding matrix ${\bf U}=[{\bf u}_1,...,{\bf u}_4]$. More specifically, ${\bf u}_1,~{\bf u}_2,~{\bf u_3},~{\bf u_4}$ are for symbols $s_{13}+s_{31}, s_{14}+s_{41}, s_{23}+s_{32}, s_{24}+s_{42}$, respectively. Each beamformer is designed as below
\begin{equation*}\label{U1}
\textbf{u}_1\subseteq \textbf{Null}
\left[
\begin{array}{ccc}
 \textbf{G}_{r,2}\\
 \textbf{G}_{r,4}
\end{array}
\right],~\textbf{u}_2\subseteq \textbf{Null}
\left[
\begin{array}{ccc}
 \textbf{G}_{r,2}\\
 \textbf{G}_{r,3}
\end{array}
\right]
\end{equation*}

\begin{equation}\label{U3}
\textbf{u}_3\subseteq \textbf{Null}
\left[
\begin{array}{ccc}
 \textbf{G}_{r,1}\\
 \textbf{G}_{r,4}
\end{array}
\right],~\textbf{u}_4\subseteq \textbf{Null}
\left[
\begin{array}{ccc}
 \textbf{G}_{r,1}\\
 \textbf{G}_{r,3}
\end{array}
\right].
\end{equation}
Since $\left[\textbf{G}_{r,j}^T,~\textbf{G}_{r,k}^T\right]^T$ is a $4\times 5$ matrix, we can always find these beamforming vectors to satisfy the above conditions.

Plugging \eqref{U3} into \eqref{y_i}, we can obtain the received signal for source node $1$ as below
\begin{eqnarray}\nonumber
{\bf y}_1={\bf G}_{r,1}{\bf U}\hat{{\bf s}}_{\oplus}+{\bf n}_1~~~~~~~~~~~~~~~~~~~~~~~~~~~~~~~~~~~\\\label{y_ii}={\bf G}_{r,1}\big[{\bf u}_1(s_{13}+s_{31})+{\bf u}_2(s_{14}+s_{41})\big]+{\bf n}_1.
\end{eqnarray}
Since source node $1$ knows $s_{13}$ and $s_{14}$, it can decode the desired signal from \eqref{y_ii} after applying self-interference cancellation. In the same manner, the other source nodes can also obtain
the signals intended for themselves. Thus, the total DoF of $8$ is achieved, which is also the upper bound for the network when $M=2, N=5$.


\subsection{Extension to $M=2$, $N>5$}
When $M=2, N>5$, we can always find the null space for $\textbf{a}_i$ (i=1,2,3,4) by \eqref{a_1}-\eqref{a_4}. The precoding matrix $\textbf{V}$ then can be calculated by \eqref{V_i}. It can be seen that $\textbf{u}_i$ exists by \eqref{U3}. Thus, we can apply our proposed GSA-based transmission scheme for MIMO two-way X relay channel to achieve the total DoF of $8$. We omit the proof here.

\section{Achievability of The Upper Bound}
In this section, we will generalize the method in the previous section to arbitrary $N$, $M$ with $M<\frac{N}{2}$ and we will show that it can achieve the DoF upper bound $4M$ when $M \leq \lfloor\frac{2N}{5}\rfloor$ for even $M$, and $M \leq \lfloor\frac{2N-1}{5}\rfloor$ for odd $M$.

We first consider the case when $M$ is even. The proposed GSA-based transmission scheme achieves total DoF upper bound $d_{13}=d_{14}=d_{23}=d_{24}=d_{31}=d_{32}=d_{41}=d_{42}=\frac{M}{2}$, yielding a total DoF of $4M$. Denote $\textbf{s}_{\oplus}=[s_{13}^1+s_{31}^1, s_{13}^2+s_{31}^2, \cdots, s_{13}^\frac{M}{2}+s_{31}^\frac{M}{2}, s_{14}^1+s_{41}^1, \cdots,  s_{14}^\frac{M}{2}+s_{41}^\frac{M}{2}, s_{23}^1+s_{32}^1, \cdots, s_{23}^\frac{M}{2}+s_{32}^\frac{M}{2}, s_{24}^1+s_{42}^1, \cdots, s_{24}^\frac{M}{2}+s_{42}^\frac{M}{2}]^T$ as the network-coded messages expected to obtain at the relay, where $s_{ij}^k$ denotes the $k$-th data streams from source node $i$ to source node $j$.

Denoting $\textbf{A}_i$ as the $(\frac{(i-1)M}{2}+1)$-th to the $(\frac{iM}{2})$-th row vectors of \textbf{A}, for $i=1, \cdots, 4$. Each ${\bf A}_i$ can be thought as a combining matrix for the transmitted signals of a source node pair. Thus, we design $\textbf{A}_i$ similar to \eqref{a_1} and \eqref{a_4} as the following:
\begin{eqnarray}\nonumber\label{a_1234}
{\bf A}_1^T\subseteq \textbf{Null}~\big[{\bf H}_{2,r}~{\bf H}_{4,r}\big]^T~\\\nonumber
{\bf A}_2^T\subseteq \textbf{Null}~\big[{\bf H}_{2,r}~{\bf H}_{3,r}\big]^T~\\\nonumber
{\bf A}_3^T\subseteq \textbf{Null}~\big[{\bf H}_{1,r}~{\bf H}_{4,r}\big]^T~\\
{\bf A}_4^T\subseteq \textbf{Null}~\big[{\bf H}_{1,r}~{\bf H}_{3,r}\big]^T.
\end{eqnarray}
 Here, $\textbf{A}_1$ is for source pair (1,3), $\textbf{A}_2$ is for source pair (1,4), $\textbf{A}_3$ is for source pair (2,3) and $\textbf{A}_4$ is for source pair (2,4).

We can see that $\textbf{A}_i^{T}$ is an $N \times \frac{M}{2}$ matrix and $\left[\textbf{H}_{j,r}~\textbf{H}_{k,r}\right]^T$ is a $2M \times N$ matrix. The matrix $\textbf{A}_i^{T}$ exists if and only if $N-2M \geq \frac{M}{2}$, which is equivalent to $M \leq \lfloor\frac{2N}{5}\rfloor$.

After obtaining the matrix \textbf{A}, we can get the matrix \textbf{V} using the same method as \eqref{V_i} in Section III. Then we show the existence of the precoding matrix \textbf{U} ($N \times 2M$). We can write \textbf{U} as
\begin{equation}\label{UU}
\textbf{U}=
\left[
\begin{array}{ccc}
 \textbf{U}_{1}~\textbf{U}_{2}~\textbf{U}_{3}~\textbf{U}_{4}
\end{array}
\right],
\end{equation}
where each $\textbf{U}_i$ is an $N \times \frac{M}{2}$ matrix and
\begin{equation*}\label{UU1}
\textbf{U}_1\subseteq \textbf{Null}
\left[
\begin{array}{ccc}
 \textbf{G}_{r,2}\\
 \textbf{G}_{r,4}
\end{array}
\right],~\textbf{U}_2\subseteq \textbf{Null}
\left[
\begin{array}{ccc}
 \textbf{G}_{r,2}\\
 \textbf{G}_{r,3}
\end{array}
\right]
\end{equation*}

\begin{equation}\label{UU3}
\textbf{U}_3\subseteq \textbf{Null}
\left[
\begin{array}{ccc}
 \textbf{G}_{r,1}\\
 \textbf{G}_{r,4}
\end{array}
\right],~\textbf{U}_4\subseteq \textbf{Null}
\left[
\begin{array}{ccc}
 \textbf{G}_{r,1}\\
 \textbf{G}_{r,3}
\end{array}
\right].
\end{equation}

We can see that $\left[ \textbf{G}_{r,j}~~\textbf{G}_{r,k} \right]^T$ is a $2M \times N$ matrix. The matrix $\textbf{U}_i$ exists if and only if $N-2M \geq \frac{M}{2}$, which is equivalent to $M \leq \lfloor\frac{2N}{5}\rfloor$. Hence, we can apply GSA-based transmission scheme when $M$ is even and $M \leq \lfloor\frac{2N}{5}\rfloor$.

Similarly, we can achieve the upper bound of the DoF $4M$ by $d_{13}=d_{24}=d_{31}=d_{42}=\frac{M+1}{2}$ and $d_{14}=d_{23}=d_{32}=d_{41}=\frac{M-1}{2}$ when $M$ is odd. The corresponding matrices \textbf{A} and \textbf{U} exist if and only if $N-2M \geq \frac{M+1}{2}$, which is equivalent to $M \leq \lfloor\frac{2N-1}{5}\rfloor$.

Note that the proposed GSA based transmission scheme can be applied to align signal pairs even when $N>2M$. In this case, $\textbf{A}$ is an identity matrix, and GSA reduces to the conventional SA.

Finally, we summarize the generalized signal alignment procedure in the following chart.

\vspace{0.4cm} \hrule \hrule \vspace{0.2cm} \textbf{Outline of Generalized Signal Alignment} \vspace{0.2cm} \hrule
\vspace{0.2cm} ~~
\begin{itemize}
\item {\bf Step 1.}~ construct the matrix \textbf{H} using the channel matrices $\textbf{H}_{1,r}$, $\textbf{H}_{2,r}$, $\textbf{H}_{3,r}$, $\textbf{H}_{4,r}$ according to \eqref{y_r_2_N}.
    \vspace{0.2cm} ~~
\item {\bf Step 2.}~ Design the relay processing matrix \textbf{A}  according to \eqref{a_1234}.
\vspace{0.2cm} ~~
\item {\bf Step 3.}~Compute the effective
channel in the MAC phase \textbf{AH} and construct the source precoding matrix \textbf{V} according to \eqref{V_i}.
\vspace{0.2cm} ~~
\item {\bf Step 4.}~ Design the BC precoding matrix \textbf{U} with matrix $\textbf{G}_{r,1}$, $\textbf{G}_{r,2}$, $\textbf{G}_{r,3}$, $\textbf{G}_{r,4}$ according to \eqref{UU3}.
\end{itemize}
\vspace{0.2cm} \hrule \vspace{0.4cm}

\section{Numerical Results}
In this section, we provide numerical results to show the sum rate performance of the proposed scheme for the MIMO two-way X relay channel. The channel between each source node and the relay node is modeled as Rayleigh distribution with unit variance and it is independent for different node. The numerical results are illustrated with the ratio of the total transmitted signal power to the noise variance at each receive antenna and the total throughput of the channel. Each result is averaged over 10000 independent channel realizations.

We now explain how we compute the sum rate for the MIMO two-way X relay channel when applying the GSA transmission scheme.

GSA transmission scheme can be used in both amplify-and-forward (AF) and decode-and-forward (DF) strategy. From \eqref{y_r_2_N} and \eqref{y_ii}, when we apply AF strategy, we can obtain
\begin{equation}\label{y_iii}
\textbf{y}_i=\textbf{G}_{r,i}\textbf{U}(\textbf{s}_{\oplus}+\textbf{A}\textbf{n}_r)+\textbf{n}_i.
\end{equation}

Let $R_i$ denote the sum rate of the source node $i$. We calculate $R_1$ as a representative. First, we write the received signal of source node 1 with AF strategy as
\begin{equation}\label{y_111}
{\bf y}_1={\bf G}_{r,1}{\bf U}\textbf{s}_{\oplus}+{\bf G}_{r,1}{\bf U}{\bf A}{\bf n}_r+{\bf n}_1~~~~~~~~~~~~~~~~~~~~~~~~~~~~~~~~~~$$$$
~~~=\underbrace{{\bf G}_{r,1}\big[{\bf U}_1(\textbf{s}_{13}+\textbf{s}_{31})+{\bf U}_2(\textbf{s}_{14}+\textbf{s}_{41})\big]}_{Signal}+\underbrace{{\bf G}_{r,1}{\textbf{U}}{\textbf{A}}{\bf n}_r+{\bf n}_1}_{Noise}$$$$
=\tilde{\textbf{G}}_{r,1}\tilde{\textbf{s}}_1+\tilde{\textbf{n}}_1.~~~~~~~~~~~~~~~~~~~~~~~~~~~~~~~~~~~~~~~~~~~~~
\end{equation}
Then we can calculate the sum rate $R_{1}$ in bits per channel use from source node 3 and source node 4 to source node 1 by \eqref{y_111}
\begin{equation}\label{R1_AF}
R_{1}=\log_2 [\det({\bf I}+
(\varepsilon(\tilde{\textbf{n}}_1\tilde{\textbf{n}}_1^H))^{-1}
\tilde{\textbf{G}}_{r,1}\varepsilon[\tilde{\textbf{s}}_1\tilde{\textbf{s}}_1^H]\tilde{\textbf{G}}_{r,1}^H)]
\end{equation}

Similarly, we can calculate $R_2$, $R_3$, $R_4$ with the same method. Then total sum rate is given by
\begin{equation}\label{R_AF}
R=\sum_{i=1}^4 {R_i}.
\end{equation}

In Fig. 4, we plot the sum rate performance of the proposed generalized signal alignment transmission scheme  at fixed $N$ but varying $M$ ($M<N/2$). In the figure, SNR denotes the total transmitted signal power from all the four source nodes to the noise variance at relay. We can observe that the increasing speed of sum-rate (the increase in bps/Hz for every 3dB in SNR) matches with the theoretical DoF $4M$ very well when SNR is high enough.

In Fig. 5 and Fig. 6, we plot the sum rate performance of the proposed GSA transmission scheme at the antenna configurations of $M = \lfloor\frac{2N}{5}\rfloor$ for even $M$ and $M = \lfloor\frac{2N-1}{5}\rfloor$ for odd $M$, respectively. The upper bound of the DoF of $4M$ is also achieved. These results indicate that our proposed GSA transmission scheme is feasible and effective.

\begin{figure}[t]
\begin{centering}
\includegraphics[scale=0.45]{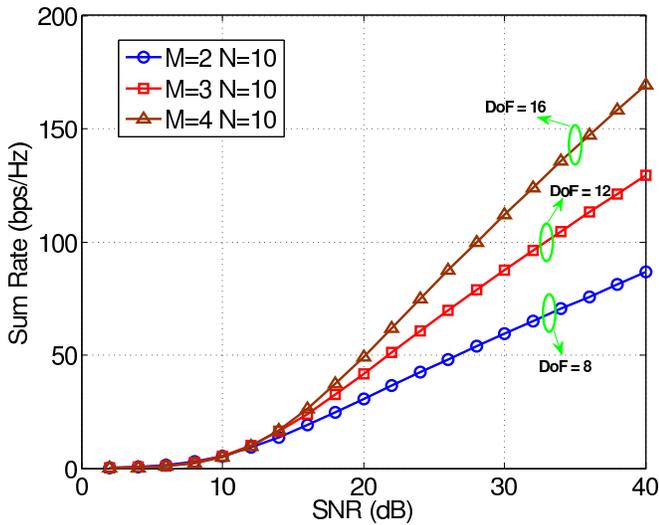}
\vspace{-0.1cm}
 \caption{Total DoF for the MIMO two-way X relay channel under generalized signal alignment transmission scheme.}
\end{centering}
\vspace{-0.3cm}
\end{figure}

\begin{figure}[t]
\begin{centering}
\includegraphics[scale=0.45]{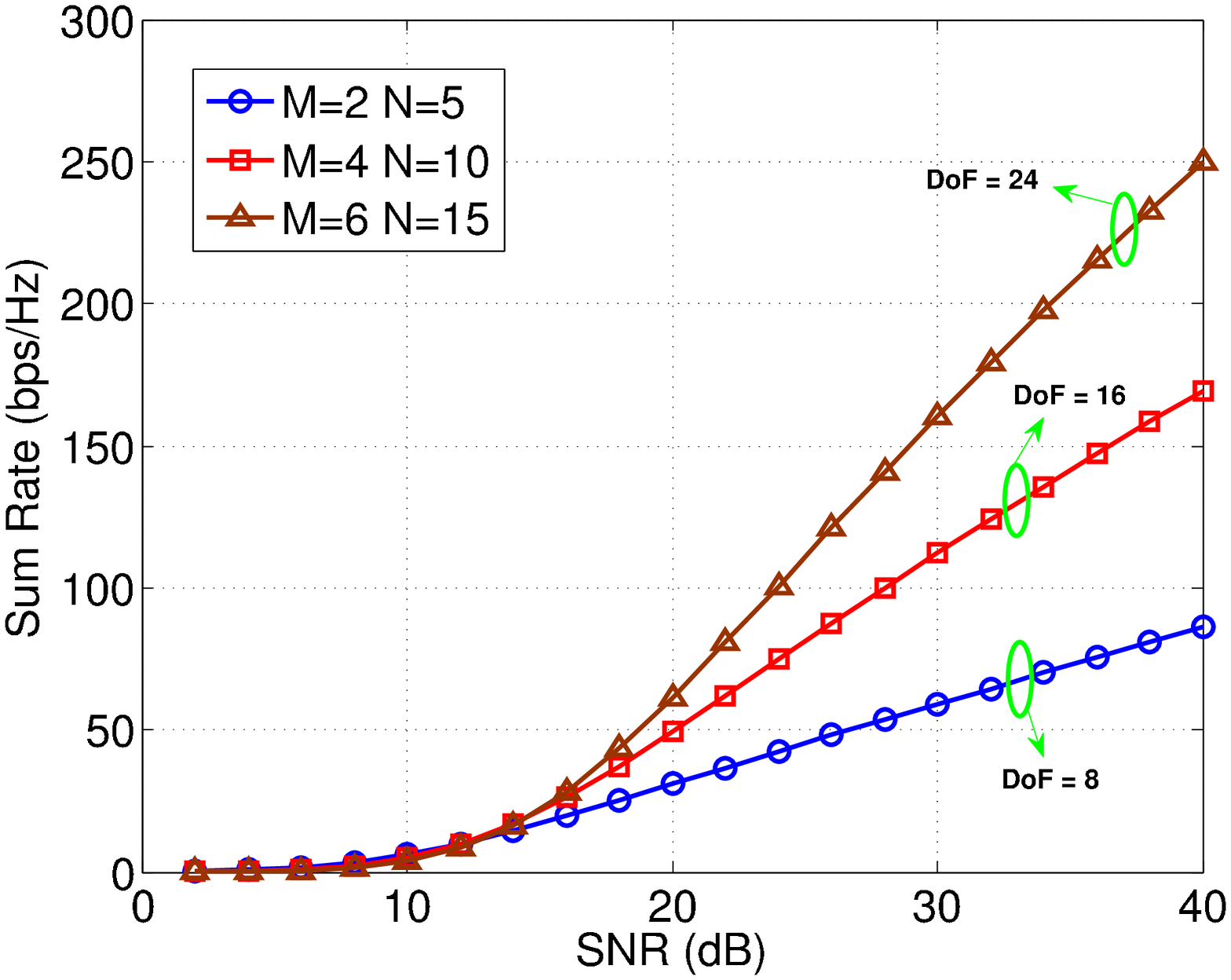}
\vspace{-0.1cm}
 \caption{Total DoF for the MIMO two-way X relay channel when $M = \frac{2N}{5}$.}
\end{centering}
\vspace{-0.3cm}
\end{figure}

\begin{figure}[t]
\begin{centering}
\includegraphics[scale=0.45]{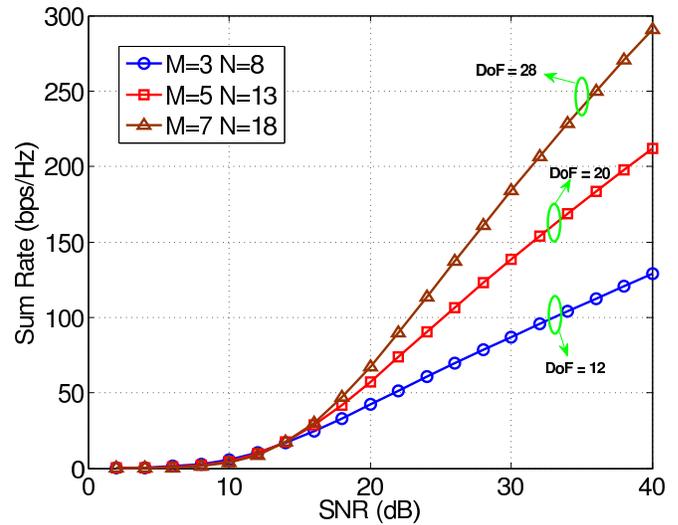}
\vspace{-0.1cm}
 \caption{Total DoF for the MIMO two-way X relay channel when $M = \frac{2N-1}{5}$.}
\end{centering}
\vspace{-0.3cm}
\end{figure}

\section{Conclusion}
In this paper, we have analyzed the achievability of the DoF upper bound for the MIMO two-way X relay channal when $M \leq \frac{N}{2}$. In the newly-proposed GSA transmission scheme, the processing matrix at the relay and the precoding matrix at the source nodes are designed jointly so that the signals to be exchanged between each source node pair are aligned at the relay. We showed that when $M \leq \lfloor\frac{2N}{5}\rfloor$($M$ is even) or $M \leq \lfloor\frac{2N-1}{5}\rfloor$($M$ is odd), the upper bound of the total DoF $4M$ is achieved. Theoretical analysis and numerical results both show that the transmission scheme proposed is feasible and effective.

\bibliographystyle{IEEEtran}
\bibliography{IEEEabrv,reference}

\end{document}